# The New ISIS Instrument Control System

F. A. Akeroyd, S. I. Campbell and C. M. Moreton-Smith
(ISIS Facility, CCLRC Rutherford Appleton Laboratory, UK)

*Data acquisition, instrument control and data analysis has traditionally been performed at ISIS using a VMS based computer system. This setup has served us well for many years, but has recently been showing signs of strain.*

*New instruments such as MAPS and GEM are generating very large data sets and the slow transfer rate of the VMS-SCSI-Data Acquisition Electrons (DAE) link (~1-2Mb/s) means ending an experiment can take a considerable amount of time. In addition new sample environment equipment often comes with an off the shelf PC driver, which means that either the VMS machine must talk to the PC over the network (two machines + communications to worry about) or a separate VMS driver must be written (development + debugging time). A decision was taken to make use of commercially available software where possible to save development time.*

*The new ISIS Engineering diffractometer instrument, ENGIN-X [1], will require short run start/stop times to perform "stroboscopic" measurements and so a new scheme has been devised. The second generation ISIS DAE (known as DAE-II) [2] is based around a VME crate and so we have chosen to make use of a commercially available PC-VME interface (MXI-2 from National Instruments [3]) which gives us a sustained data transfer rate of 23Mb/s and comes with a driver library for rapid application development. LabVIEW [4], also from National Instruments, has been chosen for sample environment control.*

*Moving away from our VMS system has also given us an opportunity to review the format used to store our RAW data. We have chosen to ultimately use the NeXus data format [5], and a draft ISIS NeXus raw data file format is currently being refined.*

## Contents

- The Old Control System
- The Need for a New System
- Choices
- Components of the New System
- Remote Access
- Data File Format
- Future Developments
- References
- Figures

## The Old Control System

Most data collection at ISIS is carried out using a VMS computer connected to locally produced Data Acquisition Electronics (DAE) via a SCSI link. The DAE processes input from the detector array and stores counts v time-of-arrival histograms in memory. If data collection is dependent on external factors, then this can be accommodated in two ways. When the external factor is slowly varying, or a fast response is not required, the VMS workstation is used: a separate process (called CAMACMON) monitors sample environment parameters (e.g. sample temperature) and can suspend/resume data collection if these move out of pre-determined ranges. For faster varying parameters, a veto signal can be fed directly into the DAE hardware - this signal is checked every ISIS pulse (0.02 seconds) and so provides a very fine degree of control. At the end of an experiment, data is transferred from the DAE to the host computer, saved to disk in a local file format (the ISIS RAW data file) and then automatically archived.

Remote access to the data collection computer is provided by telnet, and a remote user is able to start a status window (called the DASHBOARD) and issue commands to control data collection. As all the instrument and sample environment control commands are integrated into the VMS command language, a set of experiments can be run offline by creating a VMS command file (script) and using e.g. Digital Command Language (DCL) to create loops and other control structures.

### *The Need for a New System*

The old system works well and for many ISIS instruments meets all their current and future needs. Impetus for a new system has however come from several sources:

### New Sample Environment Equipment

New sample environment equipment must be incorporated into the VMS control system, which can require many days of work programming and testing. Much new equipment now comes with a ready-to-use PC control interface (we can usually get one based on LabVIEW [4]) - we would save time if this could be taken advantage of. Recently, VMS has provided support for DCOM (the standard Windows program-to-program communication protocol) so we have been using this VMS$\leftrightarrow$PC communication to make use of PC control, but this is not as stable as we would desire. Moving to a totally PC based control system would make it easier to take advantage of these off-the-shelf drivers directly and use them in a more stable environment.

### Collected Data Size

The new ISIS engineering diffractometer instrument, ENGIN-X [1], will be capable of generating very large data sets (500+ Mb) during short duration (~1 minute) experiments. Our current VMS-SCSI-DAE link is only capable of 1-2Mb/s and so not capable of meeting this requirement (you would spend more time saving your data than collecting it!). Also, we are finding that the memory to disk write performance of VMS is much slower than that of modern PCs and this delay is now becoming an issue on several instruments.

### Maintainability and Future Development

VMS is a legacy system - though its future looks assured for the next 5-10 years, developers and visiting users are less familiar with it. Tools for developing software are less widespread than for the PC, and many new analysis and visualisation packages are not made available. This means:

- Software is harder to develop, can take longer, and may not be able to use the best available technology or packages
- Users are unfamiliar with VMS and so time must be set aside to teach them how to use it
- Many instruments are already analysing their data on PC and thus have to set up systems of synchronising and moving data to and from the VMS system

Though the new ISIS DAE-2 is VME based, it can only be accessed from VMS via an additional crate SCSI interface card - this SCSI card is now obsolete and also proving to be the performance bottleneck.

### *Choices*

Given the above arguments, a PC-VME based system is the desirable step forward. As much of our sample environment came with (or already has) a LabVIEW based control interface, integrating the whole system into LabVIEW would be desirable. LabVIEW has many other advantages - it is an easy to use graphical programming language designed mainly for the purpose of data acquisition, GPIB and serial instrument control, data analysis, data presentation, and data storage. National Instruments also produce a wide range of related hardware and other software. LabVIEW:

- Reduces development time
- Has powerful analysis libraries
- Simplifies complicated programming tasks
- Is flexible enough to allow you to control any equipment
- Incorporates the latest technology, such as multi-threading and ActiveX
- Has the ability to communicate with other applications using a wide variety of protocols including: ActiveX, ftp, http, DLLs, SQL, TCP/IP sockets, National Instruments Data Sockets

- Has everything you need to write, execute, debug and distribute programs provided in one complete package
- Fully integrates with other National Instruments hardware

This then left us with the choice of Linux or Windows for the operating system. We chose Windows as, at the time, it provided more functionality and was also a much more familiar development environment for our users and staff. Another reason for choosing the National Instrument route is that they also produce a VME control environment (MXI-2), which includes interface cards for both the VME crate and PC and a library for accessing VME devices (NI-VISA). MXI-2 is capable of 32Mb/s burst and 23Mb/s sustained data transfer rates.

## *Components of the New System*

For a summary of the system layout see Fig. 1 at the end of this document.

## The Data Collection Hardware (DAE-II)

The Primary Function of the Second Generation, Data Acquisition Electronics, (DAE-II) [2], on ISIS, is to provide high capacity, high bandwidth acquisition, along with Data Histogramming, for the next generation of Neutron Instruments. The approach taken when designing the system has been to adopt a modular, off-the-shelf approach, where possible.

The Modularity enables the system to be easily expanded, according to individual instrument requirements (e.g. MAPS or SXD). This provides large amounts of Histogram Memory, along with a high number of Detector inputs - achieved through a Detector card providing both the necessary Detector Input Structure and Histogram Memory on a single board.

The System is synchronized to the ISIS Source, through the Environment Card, which is required in each Acquisition Crate used in the system. It provides Frame Synchronisation distribution and Run control to all Detector cards in the crate.

The Crate is controlled and set-up via an off-the-shelf VME Crate controller: currently this is a National Instruments MXI-2 Interface [3], although any VME Controller could be used. This enables system configuration data to be downloaded from a host PC, running the Instrument Control Program, in addition to controlling the Data Readout from the crate at the end of each Run. This Crate controller provides a stable and proven backbone from the PC to the crate, whilst enabling transfer rate up to 30 MB/s to be achieved from the Histogram Memory.

## The PC Hardware

The design for the new system centres on having two computers in the instrument cabin:

- An *instrument computer* attached to the data acquisition hardware. Normally users will not have to interact directly with this computer, but instrument scientists may need to use it to set up advanced sample environment.
- A separate *analysis/control* computer, at which users will set up scripts/command files for their experiments, look at data being collected and perform analysis.

The separation was done so that the instrument computer could be locked down into a known configuration (i.e. no random user installed or downloaded software), which provides for a more robust system; it also allows for a common configuration to be used across many instruments. Though setting up of an experiment is done on the analysis machine, the entire experiment could be controlled from the instrument computer if needed. The instrument computer contains the National Instruments PCI-MXI-2 card, along with the Windows driver and control library (NI-VISA) for accessing the VME-DAE.

## The *Ray of Light* Control Program

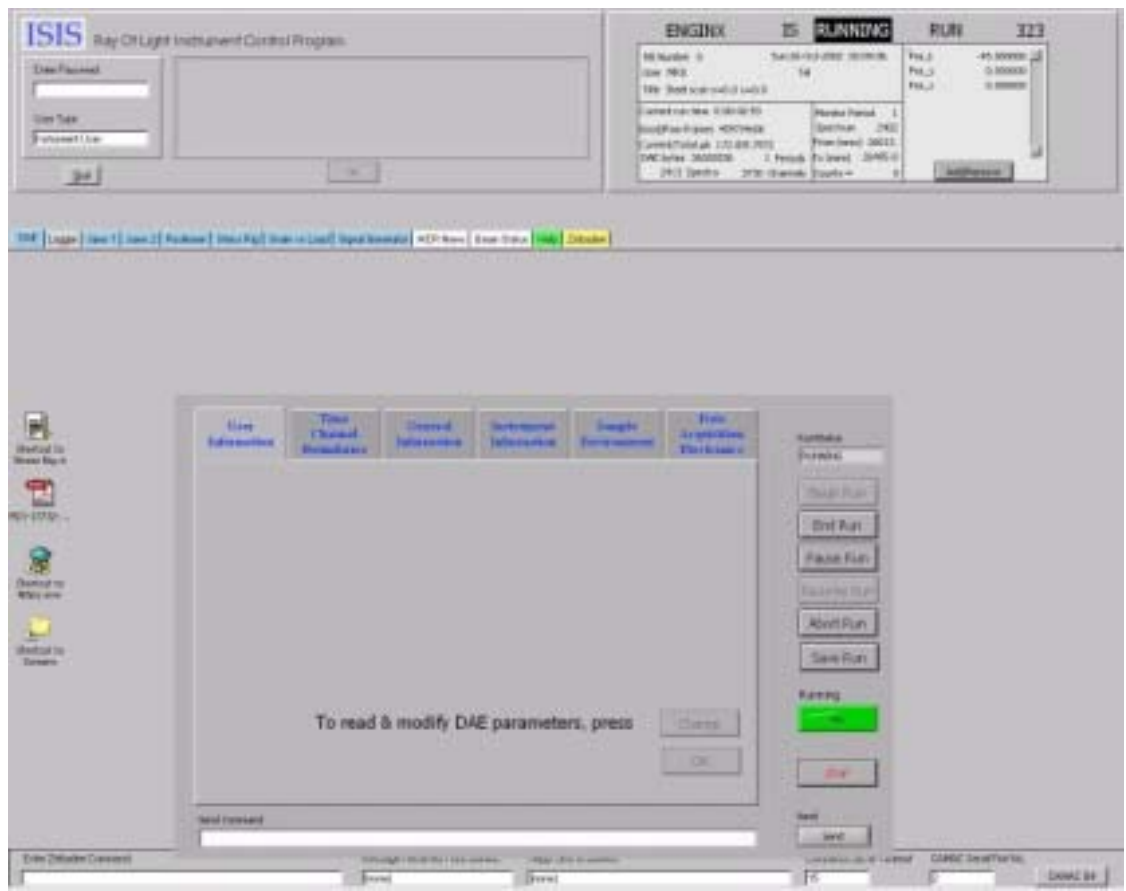

LabVIEW programs are constructed from components called VIs (Virtual Instruments). A VI is the LabVIEW equivalent of a function/procedure/subroutine in any other programming language. VIs can call other VIs and pass and return data just like functions etc. *Ray of Light* [6] is a customisable Window manager for LabVIEW VIs - it starts the appropriate ones based on a configuration file, and then arranges them into a large tabbed dialogue box for easy viewing. *Ray of Light* is written in C++ and LabWindows/CVI (LabWindows/CVI is similar to programming LabVIEW using C) - it is a framework that integrates LabVIEW VIs into a single interface.

*Ray of Light* also provides two levels of password based security access (*user* and *instrument manager*), which can be used to control which VIs are accessible and modifiable via the interface.

## The Instrument Control Program (ICP)

This program is responsible for communicating with the VME DAE, keeping track of the current run state, serving live data to the network and creating the final data file. Rather than write a completely new control program, it was decided to reuse part of the existing VMS control program to gives us the advantage of being able to produce backward compatible VMS RAW data files. This feature was very useful for early testing as it allowed us to compare data collected on the old and new systems. The SCSI access layer was replaced by a VME layer based written in C++ using the NI-VISA API. To provide remote access to the control program, we incorporated it into Open GENIE [7] as this supported automation (DCOM) access. The ICP runs as a background process and is launched by LabVIEW on system startup.

## The Dashboard VI

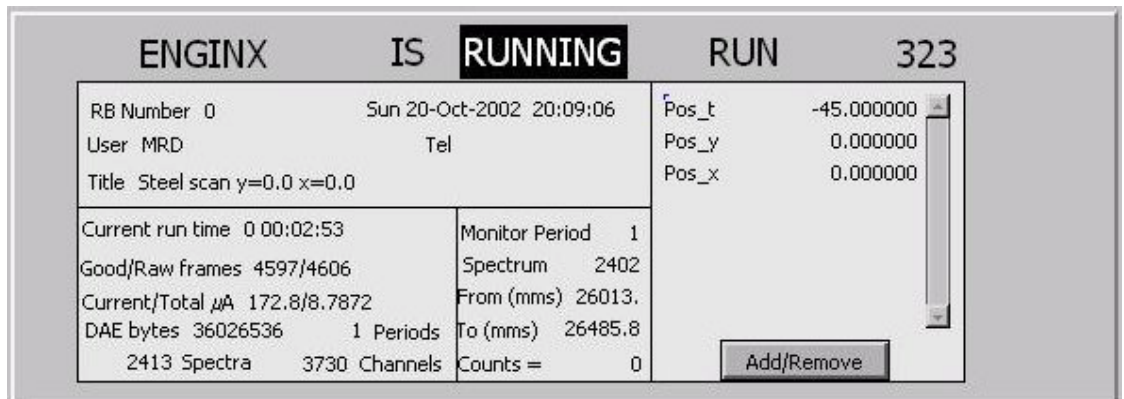

This is the instrument status panel - it is a LabVIEW VI which displays the values of important data acquisition and selected sample environment parameters. Any parameter which is added to the list of names on the right hand side of the dashboard is automatically queried every 5 seconds and logged to both a text file and the data logger process. At the end of an experiment these log files are accumulated and added to the final data file.

The names shown on the left are instrument defined *short names* for the parameters of interest - a configuration file (*controls.txt*) is used to map these names to the appropriate LabVIEW variable in the VI, which is used for setting or displaying a value.

## The Generic Data Logger VI

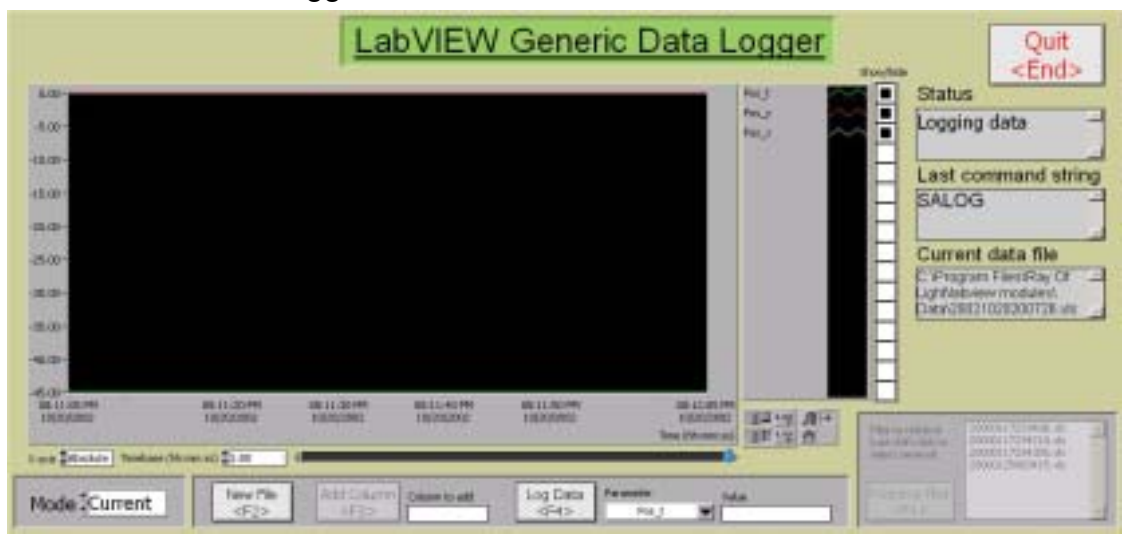

Sample environment values are automatically sent to this VI by the Dashboard, which allows them to be plotted in real time and also stored in a Microsoft Excel spread sheet table for quick viewing.

Jaws 1 | Jaws 2 | Positioner | Stress Rig | Strain vs Load | Signal Generator | MCR News | Beam Status | Help | Zebedee

Microsoft Excel - 20021020200726.xls

File  Edit  View  Insert  Format  Tools  Data  Window  Help

A1

| | A | B | C | D | E | F | G | H |
|---|---|---|---|---|---|---|---|---|
| 1 | | Pos_t | Pos_y | Pos_x | | | | |
| 2 | 20/10/2002 20:07:29 | -45 | 0 | 0 | | | | |
| 3 | 20/10/2002 20:07:34 | -45 | 0 | 0 | | | | |
| 4 | 20/10/2002 20:07:39 | -45 | 0 | 0 | | | | |
| 5 | 20/10/2002 20:07:44 | -45 | 0 | 0 | | | | |
| 6 | 20/10/2002 20:07:59 | -45 | 0 | 0 | | | | |
| 7 | 20/10/2002 20:08:04 | -45 | 0 | 0 | | | | |
| 8 | 20/10/2002 20:08:25 | -45 | 0 | 0 | | | | |
| 9 | 20/10/2002 20:08:30 | -45 | 0 | 0 | | | | |
| 10 | 20/10/2002 20:08:35 | -45 | 0 | 0 | | | | |
| 11 | 20/10/2002 20:08:40 | -45 | 0 | 0 | | | | |
| 12 | 20/10/2002 20:08:45 | -45 | 0 | 0 | | | | |
| 13 | 20/10/2002 20:08:50 | -45 | 0 | 0 | | | | |
| 14 | 20/10/2002 20:08:56 | -45 | 0 | 0 | | | | |
| 15 | 20/10/2002 20:09:01 | -45 | 0 | 0 | | | | |
| 16 | 20/10/2002 20:09:05 | -45 | 0 | 0 | | | | |
| 17 | 20/10/2002 20:09:11 | -45 | 0 | 0 | | | | |
| 18 | 20/10/2002 20:09:16 | -45 | 0 | 0 | | | | |
| 19 | 20/10/2002 20:09:21 | -45 | 0 | 0 | | | | |
| 20 | 20/10/2002 20:09:26 | -45 | 0 | 0 | | | | |
| 21 | 20/10/2002 20:09:31 | -45 | 0 | 0 | | | | |
| 22 | 20/10/2002 20:09:36 | -45 | 0 | 0 | | | | |
| 23 | 20/10/2002 20:09:41 | -45 | 0 | 0 | | | | |
| 24 | 20/10/2002 20:09:46 | -45 | 0 | 0 | | | | |
| 25 | 20/10/2002 20:09:51 | -45 | 0 | 0 | | | | |
| 26 | 20/10/2002 20:09:56 | -45 | 0 | 0 | | | | |
| 27 | 20/10/2002 20:10:01 | -45 | 0 | 0 | | | | |
| 28 | 20/10/2002 20:10:07 | -45 | 0 | 0 | | | | |
| 29 | 20/10/2002 20:10:12 | -45 | 0 | 0 | | | | |
| 30 | 20/10/2002 20:10:18 | -45 | 0 | 0 | | | | |
| 31 | 20/10/2002 20:10:24 | -45 | 0 | 0 | | | | |
| 32 | 20/10/2002 20:10:30 | -45 | 0 | 0 | | | | |
| 33 | 20/10/2002 20:10:36 | -45 | 0 | 0 | | | | |

Sheet1 / Sheet2 / Sheet3

Ready

## The Change Command VI

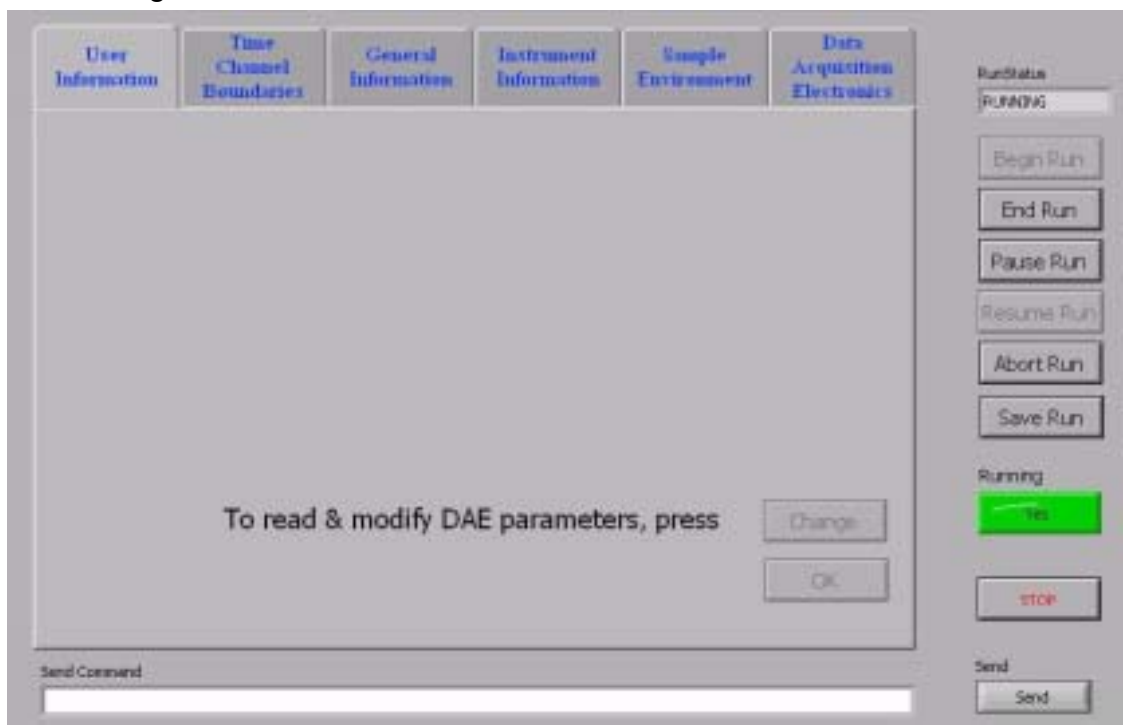

This VI provides the interface for setting up and controlling a run. From it the user can set up time channel boundaries, stop and start data collection and control spectrum and detector mapping.

## Scripting

The scientist needs the ability to be able to set up a sequence of experiments and then to let the system *get on with it*. This sequence may just involve *set temperature, collect, save, set new temperature* etc., but could also include things such as *change sample* on instruments with automatic sample changers. The old mechanism for this was to use scripting on VMS in the system supplied DCL, which works fine so long as the computer does not crash as you then loose you place in the script.

**Current Setup**

This is a system similar to that on VMS. The in-built Open GENIE Command Language (GCL) [8] was chosen as the scripting language as its syntax was flexible and well liked. It also supports DCOM calls and so can communicate easily with a LabVIEW process either on the same machine or over a Network. GCL commands are provided for *setting*, *reading* and *waiting for* values and any other Open GENIE control construct or powerful analysis feature can also be used.

As far as possible, the remote computer interacts directly with the LabVIEW front panel, pushing buttons and reading indicators. Sample environment parameters are referred to by the short name set up on the Dashboard, which is then mapped to the correct VI variable using the *controls.txt* file mentioned above. Low level commands such as *PuchLabviewButton*, *SetLabviewVariable* and *ReadLabviewVariable* are defined and used by higher level commands (such as MOTOR and PT in the example below). A few circumstances, however, require it to interact directly with Open GENIE ICP, such as reading data from the DAE - this is again done using DCOM. An example of a script is the following - PT and MOTOR are instrument specific GENIE functions for controlling sample environment equipment.

```
LOCAL iy yloc locpt    # define variables to be used
LOOP iy FROM 1 TO 11
 yloc=-2.0*iy+12.0
 print "yloc="+as_string(yloc)
 motor/y yloc
 waitfor seconds=5
 locpt=pt()
 waitfor seconds=1
 locpt=pt()
```

```
 print "pt.y="+as_string(locpt.y)
 BEGIN
   change title="Steel scan x="+as_string(locpt.y)
   waitfor seconds=5
   waitfor uamps=1
 END
ENDLOOP
```

**Planned Final Configuration**

The ultimate goal is to provide a *table driven* approach to instrument control - that is a table of separate command steps that would determine at any point what state the system was in and what conditions should be met to move onto the next step. Each step (or line of the table) would be *stateless* i.e. not depend on the action of the previous step and thus contain enough information to be self-contained. If the computer were to crash, the experiment could be restarted at the last step (unless maybe the step was marked as non-restartable). The plan is for this table to be initially produced by the script commands described above and then submitted to the control instrument PC for executing, but work on this is at an early stage.

## *Remote Access*

Experiments often need to be checked on, and adjusted, from afar. Currently we make use of 3 mechanisms:

## VNC

VNC [9] allows a remote user to access the desktop of a PC and control it as if they were sat there. It is very convenient to use, but requires a good fast network connection. As both the local and remote users have simultaneous access, it is a good tool for an instrument scientist to remotely show a local visitor how to do something.

## Windows XP Remote Desktop

This allows access similar to VNC, but is better on slower networks. The main difference is that when the remote user is connected, any local user cannot see or access the desktop.

## LabVIEW web server

This provides a mechanism to put any LabVIEW VI into a web page, thus allowing remote status information to be displayed. The latest version of LabVIEW will also allow setting of control values via the Web, but we have not yet investigated using this.

## *Data File Format*

We are looking to use the NeXus file format [5] for our RAW data files. NeXus provides a mechanism for writing the data, but the structure of the file needs to be defined separately. We are currently working on a draft ISIS NeXus file format and will use Open GENIE to produce the file.

## *Future Developments*

Information entering the DAE is in the form of a descriptor made up of two parts: a *time* part indicating when the event happened, and a *position* part indication the detector on which the event occurred. A new feature of DAE-II is a descriptor mapping table and a special extra area of memory called *focussing memory*. The descriptor mapping table is of size (number of time bins x number of positions) and is used to map an event into a spectrum in focussing memory. By correctly setting up this table, it is possible to perform real time focussing of spectra within the DAE itself. It was originally envisioned this mechanism would be used to spy on data collection statistics, with proper software analysis being done later on the full data. However, interest has been expressed in using these focussed spectra directly as they could be unloaded very quickly and so run turnaround would be quick.

## *References*

1. The ISIS ENGIN-X Engineering Diffractometer instrument, http://www.isis.rl.ac.uk/engineering/
2. ISIS Second Generation Data Acquisition Electronics (DAE-II) - for information contact J.Norris@rl.ac.uk

3. The MXI-2 interface from National Instruments, http://www.ni.com/
4. The LabVIEW package from National Instruments, http://www.ni.com/labview/
5. The NeXus data file format, http://www.neutron.anl.gov/NeXus/
6. The Ray of Light Instrument Control Program, C. M. Moreton-Smith and A. M. Rice, http://www.isis.rl.ac.uk/computing/renaissance/index.htm
7. *Open GENIE Reference Manual*, Technical Report RAL-TR-1999-031, Rutherford Appleton Laboratory, UK - http://www.isis.rl.ac.uk/OpenGENIE/
8. *Open GENIE - Analysis and Control*, arXiv paper cond-mat/0210442, NOBUGS2002/025 - see http://www.isis.rl.ac.uk/NOBUGS2002/
9. VNC Virtual Network Computing software, A T & T Research Labs, Cambridge - http://www.uk.research.att.com/vnc/

Analysis/Control Computer                    Instrument Computer

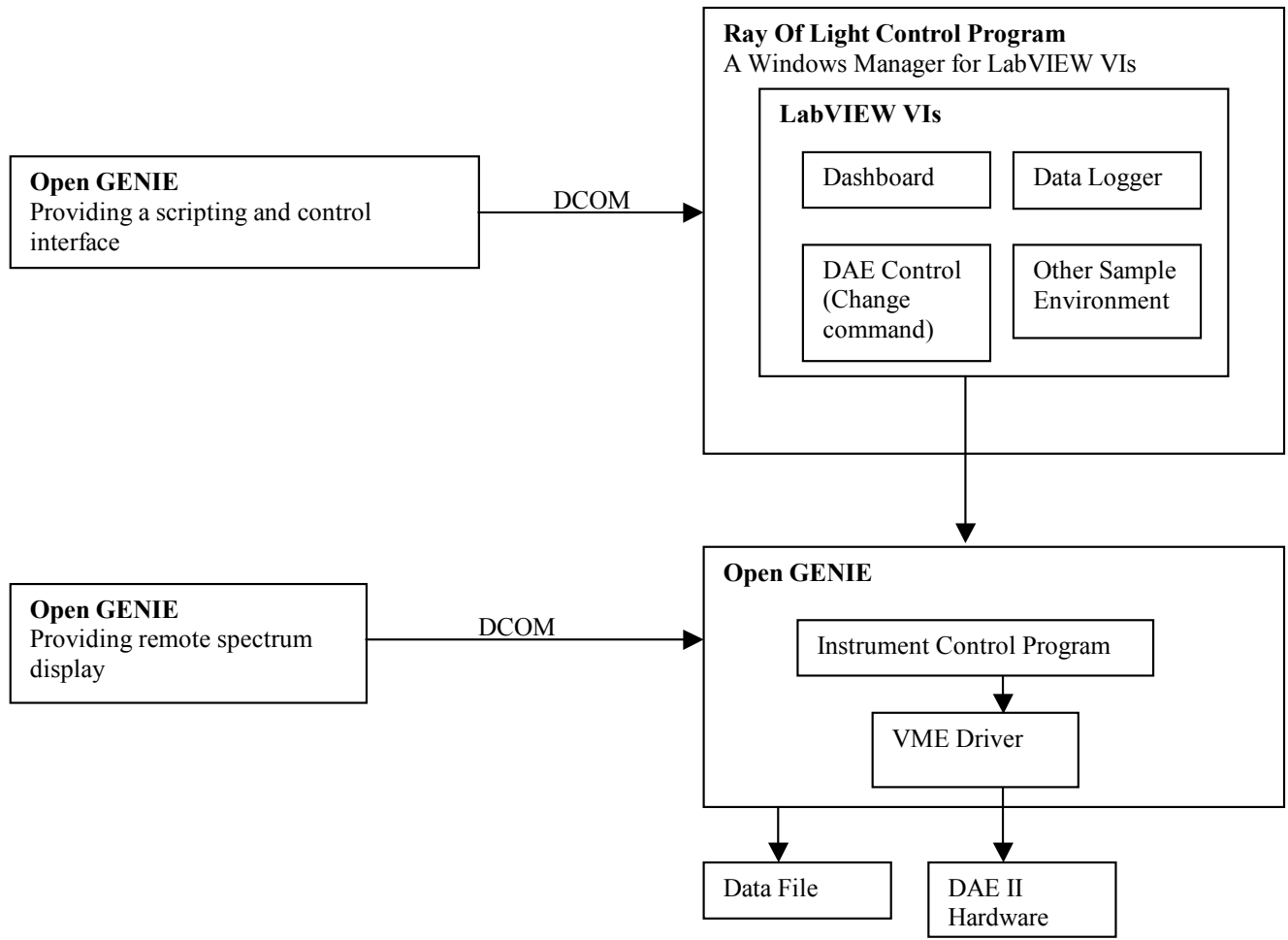

**Fig. 1** Components of the New ISIS Instrument Control System